\documentclass[10pt, letterpaper, notitlepage, oneside]{article}

\usepackage[
	letterpaper,
	bottom      = 1.00in,
	left        = 0.75in,
	right       = 0.75in,
	top         = 1.00in
]{geometry}
\usepackage[utf8]{inputenc}
\usepackage[nopar]{lipsum}

\usepackage{fancyhdr}
\fancypagestyle{specialfooter}{
	\fancyhf{}
	\lhead{Published in the \textit{Journal of Biomechanics}: \href{https://doi.org/10.1016/j.jbiomech.2018.07.023}{\url{https://doi.org/10.1016/j.jbiomech.2018.07.023}} \hfill \thepage}
	\cfoot{$\dag$ \textbf{Email address for correspondence:} g\_dilabb{\MVAt}encs.concordia.ca}
}

\usepackage{hyperref}
\hypersetup{
	breaklinks = true,
	citecolor  = blue,
	colorlinks = true,
	filecolor  = blue,
	linkcolor  = blue,      
	urlcolor   = blue
}

\usepackage{titlesec}
\titleformat*{\section}{\Large\bfseries}
\titleformat*{\subsection}{\normalsize\bfseries\filcenter}
\titleformat*{\subsubsection}{\normalsize\bfseries}
\renewcommand{\thesection}{\Roman{section}} 

\makeatletter
\renewcommand\@seccntformat[1]{\csname the#1\endcsname.\quad}
\makeatother

\usepackage{caption}
\usepackage{cleveref}
\usepackage{graphicx}
\usepackage{epstopdf}
\usepackage[
	font = normalsize
]{subfig}
\usepackage{xparse}
\crefformat{subfigure}{#2\onlyletter{#1}#3}
\crefrangeformat{subfigure}{#1--#3\onlyletter{#2}#4}
\ExplSyntaxOn
\NewDocumentCommand{\onlyletter}{m}{
	\tl_set:Nx  \l_tmpa_tl { #1 }
	\tl_item:Nn \l_tmpa_tl { -1 }
}
\ExplSyntaxOff

\usepackage{amsfonts}
\usepackage{amsmath}
\usepackage{marvosym}
\usepackage{textcomp}
\usepackage{upgreek}
\newcommand{\volume}{{\ooalign{\hfil$V$\hfil\cr\kern0.08em--\hfil\cr}}}

\usepackage{arydshln} 
\usepackage{multirow}
\usepackage{setspace}

\usepackage{apacite} 

\newcommand{\noop}[1]{}

\begin{document}
	\thispagestyle{specialfooter}
	
	\noindent
	\textbf{\LARGE Jet collisions and vortex reversal in the human left ventricle} \hfill\break
	
	\begin{changemargin}{0.25in}{0.00in}
		{\large Di Labbio G$^\dag$, Kadem L} \hfill\break
		\textit{Laboratory of Cardiovascular Fluid Dynamics, Concordia University, Montr\'{e}al, QC, Canada, H3G 1M8} \hfill\break
		
		Unnatural dynamics of the notorious vortex in the left ventricle is often associated with cardiac disease. Understanding how different cardiac diseases alter the flow physics in the left ventricle may therefore provide a powerful tool for disease detection. In this work, the fluid dynamics in the left ventricle subject to different severities of aortic regurgitation is experimentally investigated by performing time-resolved particle image velocimetry in a left heart duplicator. Diastolic vortex reversal was observed in the left ventricle accompanied by an increase in viscous energy dissipation. Vortex dynamics and energy dissipation may provide useful insights on sub-optimal flow patterns in the left ventricle.
		\hfill\break
		
		\hfill\break
		* \textit{Data pertaining to this article will be made available from the authors upon reasonable request.}\hfill\break
		* \textit{The authors have no conflicts of interest to declare.}\hfill\break
		* \textit{Please cite as}: Di Labbio, G., \& Kadem, L. (2018). Jet collisions and vortex reversal in the human left ventricle. \textit{Journal of Biomechanics}, \textit{78}, 155-160.
		
		\hfill\break
		\raisebox{0.95pt}{\small\textcopyright} 2018 Elsevier. This manuscript version is made available under the CC BY-NC-ND 4.0 license, more information regarding usage terms can be found at \href{http://creativecommons.org/licenses/by-nc-nd/4.0/}{\url{http://creativecommons.org/licenses/by-nc-nd/4.0/}}. The published version of the manuscript is available at \href{https://doi.org/10.1016/j.jbiomech.2018.07.023}{\url{https://doi.org/10.1016/j.jbiomech.2018.07.023}}. The supplementary information associated with this manuscript can be found along with the published article at the same link.
	\end{changemargin}
	
	\section{\label{sec:Intro}Introduction}
		
		The left ventricle (LV) is the heart's powerhouse, responsible for supplying each and every tissue throughout the body with oxygenated blood. During its filling phase (diastole), a vortical structure is known to develop in the LV, which has inspired a number of fluid dynamic studies over the past two decades owing to its potential for early detection of disease. This vortex facilitates the ejection of blood (systole) by redirecting the eccentric inflow toward the LV outflow tract \cite{Kilner00} with some evidence that it also minimizes energy dissipation \cite{Pedrizzetti05} and blood residence time \cite{Hendabadi13}. Unnatural dynamics of this vortex may be correlated with disease \cite{Narula07}, thus inspiring methods to characterize the behavior of a ``healthy'' vortex such as verifying the energy dissipation characteristics of the flow \cite{Pedrizzetti05, Stugaard15, Raymondet16, DiLabbio16} and defining a left ventricular vortex formation time \shortcite{Gharib06}. This work focuses on the dynamics of this notorious vortex in the presence of a particularly prevalent valvular disease, namely aortic regurgitation.
		
\begin{figure}[!h]
			\centering
			\includegraphics[width=0.70\textwidth]{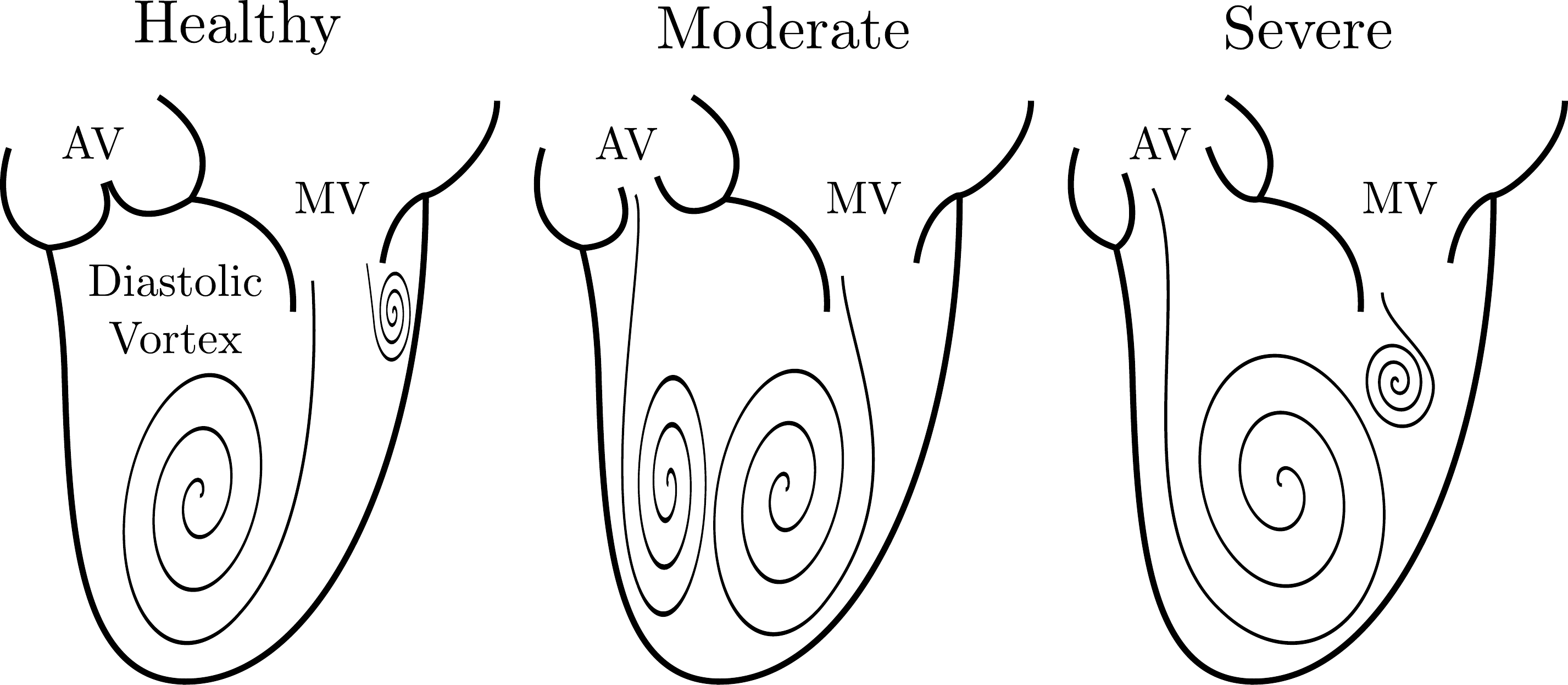}
			\caption{Schematic of the interaction between a regurgitant jet emanating from the aortic valve (AV) and the diastolic vortex developing from ordinary mitral valve (MV) inflow as the regurgitation worsens based on the results of this study (see Sec.\ \ref{sec:ResDisc} for more detail).}
			\label{fig:ARschem}
		\end{figure}

		Aortic regurgitation (AR) is a condition where the aortic valve does not adequately close. In such a case, LV filling partly occurs from leakage through the aortic valve, leading to a disruption of the formation of the diastolic vortex due to the interaction between the mitral inflow and the emergent aortic regurgitant jet (Fig.\ \ref{fig:ARschem}). AR itself is quite common, with mild AR or worse having an estimated prevalence of 5.2\% in the adult population over age 40 in the USA \cite{Singh99}. Of particular interest here is the case of chronic AR, where the left ventricle gradually responds to the persistent regurgitation by dilating in an attempt to accommodate the regurgitant volume and maintain a healthy forward stroke volume, while the peak aortic pressure may remain relatively unchanged or increase \cite{Bekeredjian05, Stout09}.
		
	\section{\label{sec:Methods}Methods}
		
\begin{figure}[!h]
			\centering
			\includegraphics[width=0.65\textwidth]{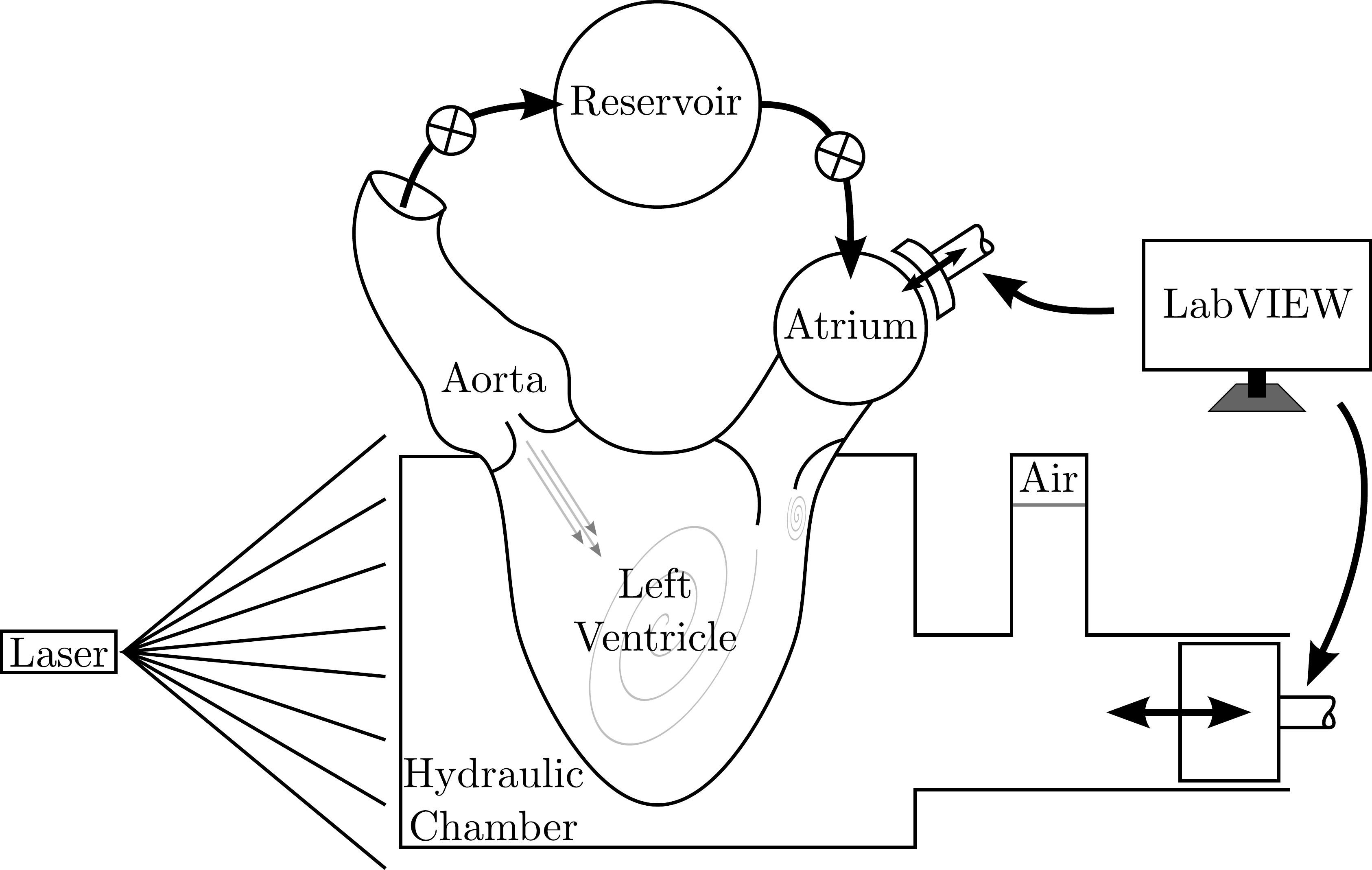}
			\caption{Schematic of the left heart duplicator in the plane of the laser sheet (i.e.\ the view of the camera). The piston acting on the hydraulic chamber controls ventricular systole (ejection) and diastole (filling). Atrial diastole occurs passively from the reservoir, while atrial systole is controlled by direct atrial compression. This independent activation of the left ventricle and left atrium during diastole (double-activation) is critical in simulating aortic regurgitation.}
			\label{fig:Dupschem}
		\end{figure}

		The intraventricular flow in the context of chronic AR was studied by experiment in a double-activation left heart duplicator (Fig.\ \ref{fig:Dupschem}). The duplicator consists of elastic silicone (SILASTIC RTV-4234-T4, The Dow Chemical Company; USA) models of the left atrium, left ventricle and aorta, having refractive indices of $1.41 \pm 0.01$. The ventricle (model made available in Appendix \ref{app:Suppl}) is encased in a hydraulic chamber with its activation being controlled by a piston-cylinder-type assembly. The left atrium is directly connected to an elevated reservoir, allowing atrial filling to occur passively. LV filling occurs in two distinct phases: ventricular relaxation (\textit{E} wave) and atrial contraction (\textit{A} wave). At the start of filling, the hydraulic piston pulls back, expanding the ventricle as it fills through the mitral valve ($23$ mm Perimount Magna Ease, Edwards Lifesciences; USA). As the piston attains its maximal backward stroke, a servomotor compresses the atrium to provide the \textit{A} wave. This begins the piston's forward stroke, hydraulically compressing the ventricle to enable ejection through the aortic valve ($25$ mm Perimount Theon RSR, Edwards Lifesciences; USA) up to the reservoir, completing the cardiac cycle. The reader is referred to the supplementary information contained in Appendix \hyperref[app:Suppl]{\ref*{app:Suppl}.1} for further technical information regarding the system. The working fluid was a mixture of $60$\% water and $40$\% glycerol by volume, having a refractive index of $1.39$ and giving a measured dynamic viscosity ($4.2$ cP) and density ($1100$ kg/m$^3$) relatively similar to those of blood.
		
		The term ``double-activation'' refers to independent activation of the LV and left atrium to respectively produce the \textit{E} and \textit{A} waves of filling. This is critical in experimentally simulating AR, as most left heart duplicators produce the \textit{A} wave by providing an additional ventricular expansion using the hydraulic piston, which would induce prolonged, non-realistic, regurgitation.
		
		AR was introduced by pulling apart graduated rods hooked under each aortic valve leaflet to restrict valve closure to a centralized orifice while allowing the leaflets to open freely (refer to Fig.\ \ref{fig:Regschem} for a schematic and to the supplementary information in Appendix \hyperref[app:Suppl]{\ref*{app:Suppl}.1} for more details). Five different severities and one ``healthy'' scenario were investigated by changing the regurgitant orifice area ($\mathrm{ROA}$) to be $0$ (normal), $0.10$ (mild), $0.18$ (moderate-1), $0.25$ (moderate-2), $0.52$ (severe-1) and $0.78$ cm$^2$ (severe-2), in accordance with clinical guidelines \cite{Nishimura14}. These orifices respectively represent $0$, $3.3$, $5.9$, $8.5$, $17.2$ and $26.1$\% of the fully open geometric aortic valve area ($3.00$ cm$^2$). The physical requirement of maintaining a constant forward stroke is equivalent to increasing the total stroke such that the volume contributed from mitral inflow remains constant. As such, the mitral inflow volume was held constant at $64 \pm 4$ mL, with the \textit{A} wave corresponding to $21 \pm 4$ mL, at a heart rate of $70$ bpm ($T = 0.857$ s), while the peak aortic pressure was maintained at $121 \pm 5$ mmHg. A normalized time, $t^* = t/T$, is here defined with $t^* = 0$ marking the start of the ejection phase, $t^* = 0.44$ the beginning of the filling phase (end of ejection), and $t^* = 1$ the end of the cycle.
		
		\begin{figure}[!b]
			\centering
			\includegraphics[width=0.65\textwidth]{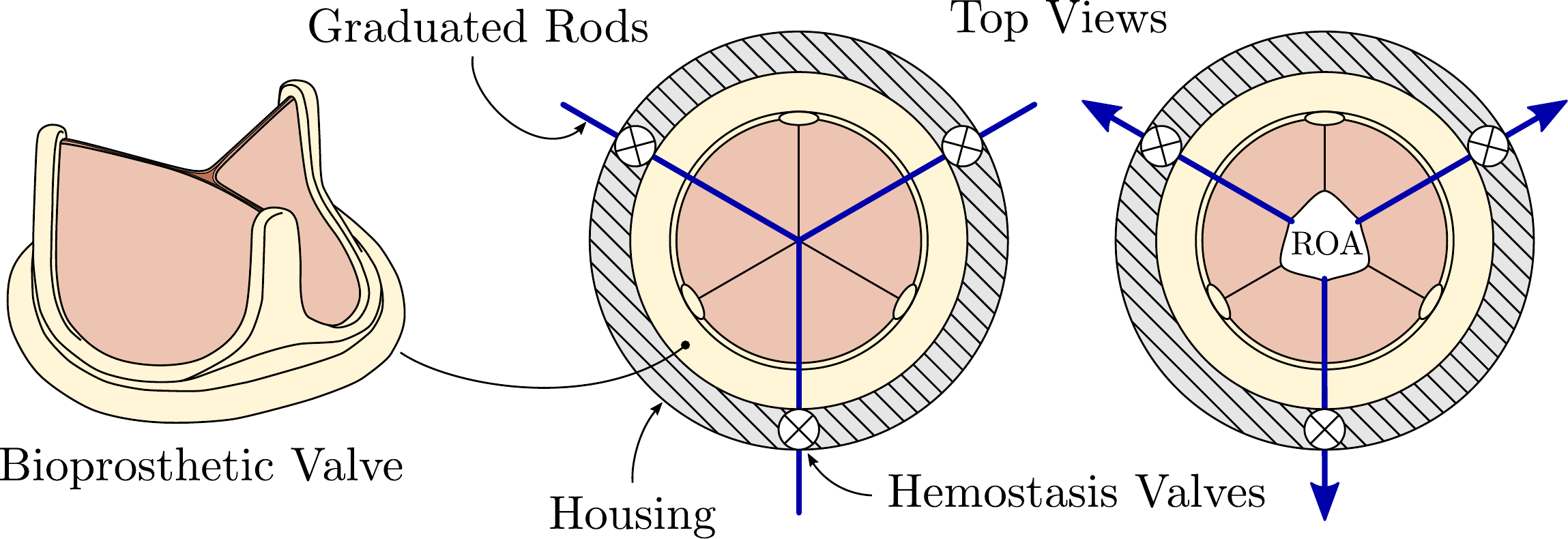}
			\caption{Schematic of the mechanism used to impose aortic regurgitation in the cardiac duplicator. Graduated rods hooked under each aortic valve leaflet were pulled apart to maintain a centralized regurgitant orifice area ($\mathrm{ROA}$) for five cases. The rods were removed to simulate the healthy flow scenario.}
			\label{fig:Regschem}
		\end{figure}
		
		The flow circuit was seeded with polyamide particles (mean diameter: $50$ $\upmu$m, density: $1030$ kg/m$^3$), to capture the velocity field by 2D time-resolved particle image velocimetry. The particles were illuminated using a double-pulsed Nd:YLF laser (LDY301, Litron Lasers; England) forming a $1$ mm thick laser sheet at the test section. The LV flow domain was captured at a recording rate of $400$ Hz (double-frame images $700$ $\upmu$s apart) at the full resolution of $1632 \times 1200$ pixels using a CMOS camera (Phantom v9.1, Vision Research Inc.; USA). The flow was recorded over one cycle in the plane crossing both the mitral and aortic valves and the ventricle apex, a view often used in clinical practice to evaluate AR \cite{LancellottiTribouilloy13}. The velocity fields were computed in DaVis 8.2 (LaVision GmbH; Germany) using a multi-pass cross-correlation algorithm with decreasing window size ($64 \times 64$ pixels down to $16 \times 16$ pixels, using $50$\% overlap). The final spatial resolution was $0.52 \times 0.52$ mm. The total uncertainty in the velocity field was estimated to be below $5$\% with respect to the maximum pointwise velocities observed in the healthy ($1.70$ m/s) and severe ($1.66$ m/s) cases, based on the major uncertainties described in \shortciteA{Raffel07} and \citeA{Adrian11}.
		
	\section{\label{sec:ResDisc}Results \& Discussion}
		
		\subsection{\label{sec:GenFlow}Replication of healthy left ventricular flow}
			
			The experiment replicates the healthy intraventricular flow rather well. A vortex ring rolls up from the shear layer of the mitral inflow, one side of which dissipates against the ventricular wall as the other side imparts a swirling motion to the entire ventricular volume \cite{Pedrizzetti05}; see Fig.\ \ref{fig:HealthyFlow} and Video 1 in Appendix \hyperref[app:Suppl]{\ref*{app:Suppl}.2}. Virtual particles released at the start of diastole exhibit a clockwise swirl and are readily aligned for ejection by the start of systole (Fig.\ \ref{fig:HealthyFlow}, right panel). The dimensionless vortex formation time \cite{Gharib06} is calculated to be $4.1$, falling in the ``healthy'' range.
			
			\begin{figure}[!h]
				\centering
				\includegraphics[width=0.65\textwidth]{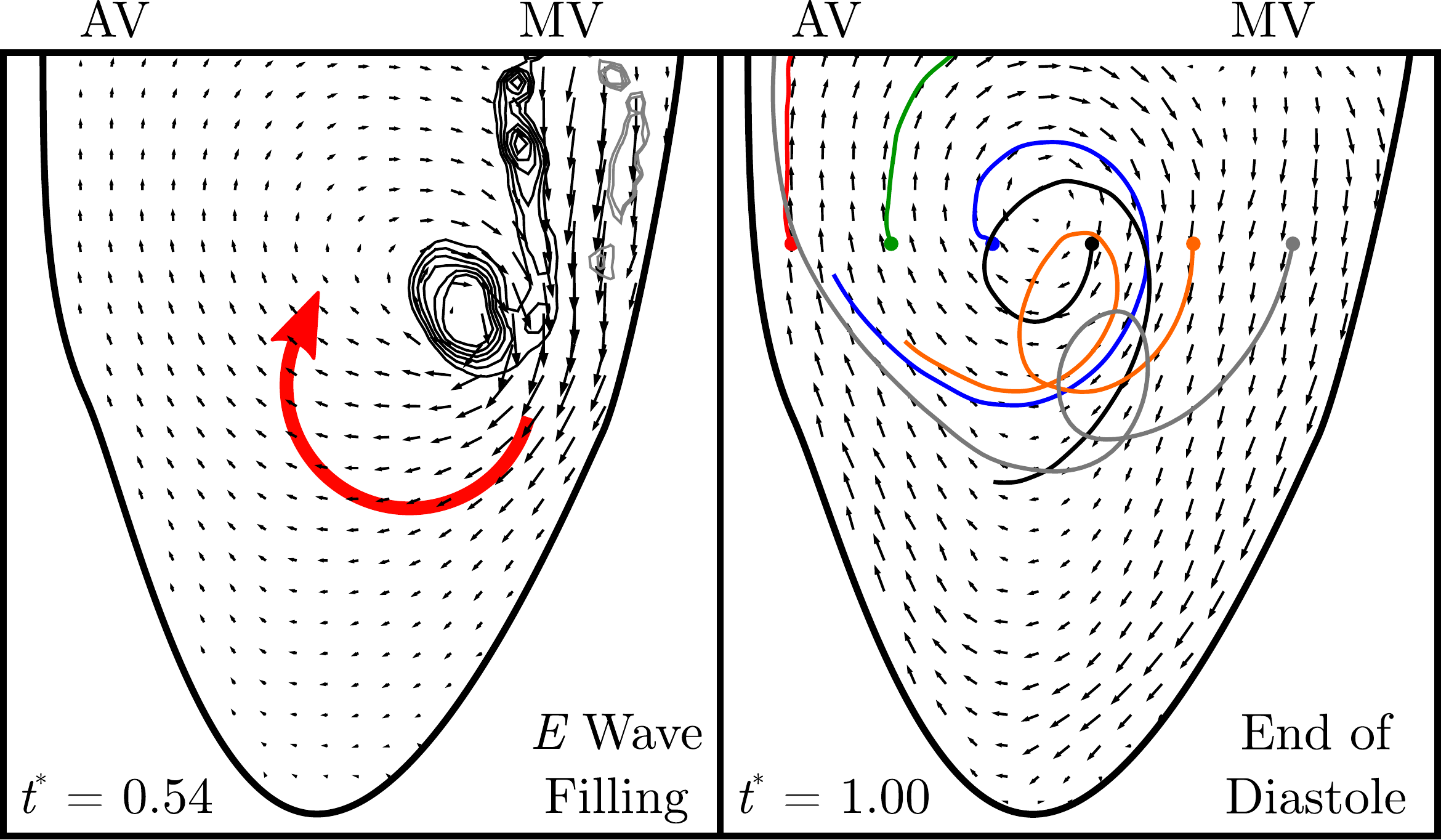}
				\caption{Healthy intraventricular flow. Left: Shear layer roll-up during \textit{E} wave filling (relaxation of the left ventricle). Isolines of vorticity are shown for selected values between $-350$ and $350$ s$^{-1}$ with negative (clockwise) vorticity in black and positive (counter-clockwise) vorticity in gray. Right: Complete clockwise swirling motion is attained at the end of diastole. Six particle pathlines released at the start of diastole ($t^* = 0.44$) exhibit a single swirling motion and are readily oriented for ejection. AV: Aortic Valve, MV: Mitral Valve.}
				\label{fig:HealthyFlow}
			\end{figure}
			
		\subsection{\label{sec:VortexRev}Vortex reversal and flow behavior}
		
\begin{figure}[!t]
				\centering
				\includegraphics[width=0.65\textwidth]{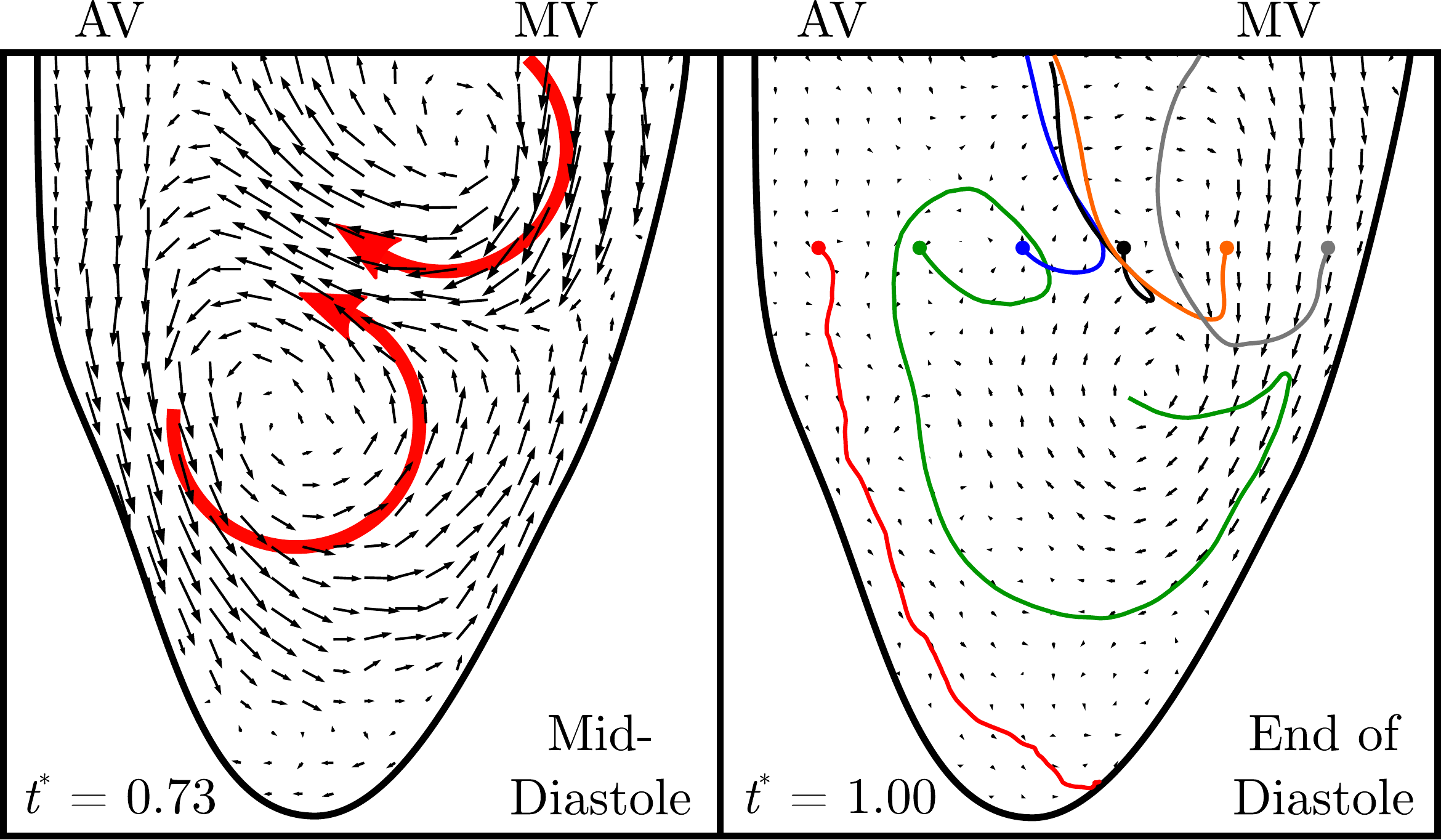}
				\caption{Intraventricular flow following severe regurgitation ($26.1$\% $\mathrm{ROA}$). Left: Demonstration of reversed vortical flow appearing in severe aortic regurgitation at mid-diastole. Right: The clockwise swirling motion of particles is lost at the end of diastole. Six particle pathlines released at the start of diastole ($t^* = 0.44$), from the same positions as in Fig.\ \ref{fig:HealthyFlow}, do not align favorably for ejection. AV: Aortic Valve, MV: Mitral Valve.}
				\label{fig:SevereFlow}
			\end{figure}

\begin{figure}[!h]
				\centering
				\includegraphics[width=0.65\textwidth]{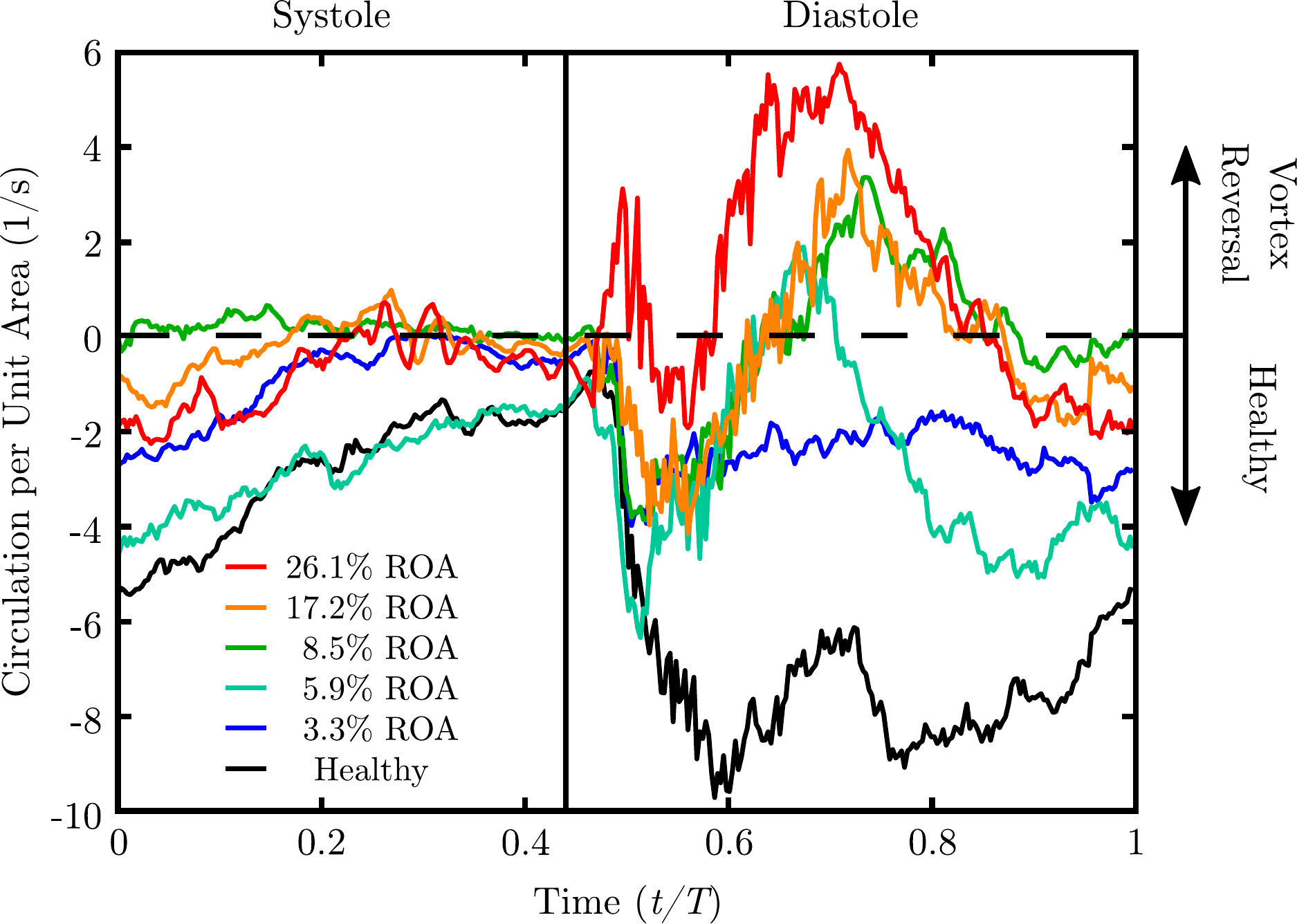}
				\caption{The circulation per unit area (or spatial mean vorticity) progressively tends toward positive values with regurgitation severity, associated with a general counter-clockwise sense of rotation.}
				\label{fig:MVORT}
			\end{figure}

			As the AR severity worsens, the regurgitant jet readily interacts with the mitral inflow. The two jets remain rather confined to the ventricle walls and each generate their respective vortices as their shear layers roll up. The two vortices effectively compete for space in the LV, with the regurgitant vortex gradually gaining the upper hand and occupying more of the ventricular volume as the severity worsens. Initially, in the mild and moderate-1 cases ($\mathrm{ROA} = 3.3$ and $5.9$\%), the regurgitant jet emanates after the mitral inflow. In the former, the healthy vortical flow does not seem to be significantly disturbed as the small regurgitant volume is simply entrained by the mitral vortex generated during the \textit{E} wave, limiting its progression into the ventricle. However, filling from the mitral valve is noticeably less forceful, having diminished velocity magnitudes, and the vortex core is ultimately unable to set up in the ventricle's center. In the latter, when the regurgitant jet emanates, it pushes the mitral vortex toward the wall, allowing the regurgitant jet to penetrate deeper into the ventricle and establish a counter-rotating vortex resembling the middle schematic of Fig.\ \ref{fig:ARschem}. In the moderate-2 case ($\mathrm{ROA} = 8.5$\%), the regurgitant jet and the mitral jet begin nearly at the same time; Video 2 in Appendix \hyperref[app:Suppl]{\ref*{app:Suppl}.3}. With the mitral inflow volume exceeding the regurgitant volume, the mitral jet sets up the more dominant vortex. The subsequent flow in this scenario is rather interesting as it represents a critical condition dependent on the timing of the two emerging jets, resembling either the moderate-1 case (Fig.\ \ref{fig:ARschem}, middle schematic) or the severe cases (Fig.\ \ref{fig:ARschem}, rightmost schematic). In the severe cases ($\mathrm{ROA} = 17.2$ and $26.1$\%), the regurgitant jet precedes the mitral inflow and establishes a counter-rotating vortex occupying the base of the ventricle; Video 3 in Appendix \hyperref[app:Suppl]{\ref*{app:Suppl}.4}. This is demonstrated in Fig.\ \ref{fig:SevereFlow} (left panel), where the regurgitant vortex imparts a completely reversed swirling motion to the flow at the base of the ventricle by mid-diastole and restricts further progression of the mitral inflow. In this case, particles released at the start of diastole ($t^* = 0.44$) are not favorably oriented for ejection, some of which follow a reversed swirling path compared to the healthy scenario (Fig.\ \ref{fig:SevereFlow}, right panel). To illustrate this vortex reversal further, in Fig.\ \ref{fig:MVORT}, the temporal evolution of the circulation in the ventricle per unit area is plotted, showing a general progression toward positive values (counter-clockwise rotation) with regurgitation severity.
			
		\subsection{Energy loss characteristics}
			
\begin{figure}[!b]
				\centering
				\includegraphics[width=0.65\textwidth]{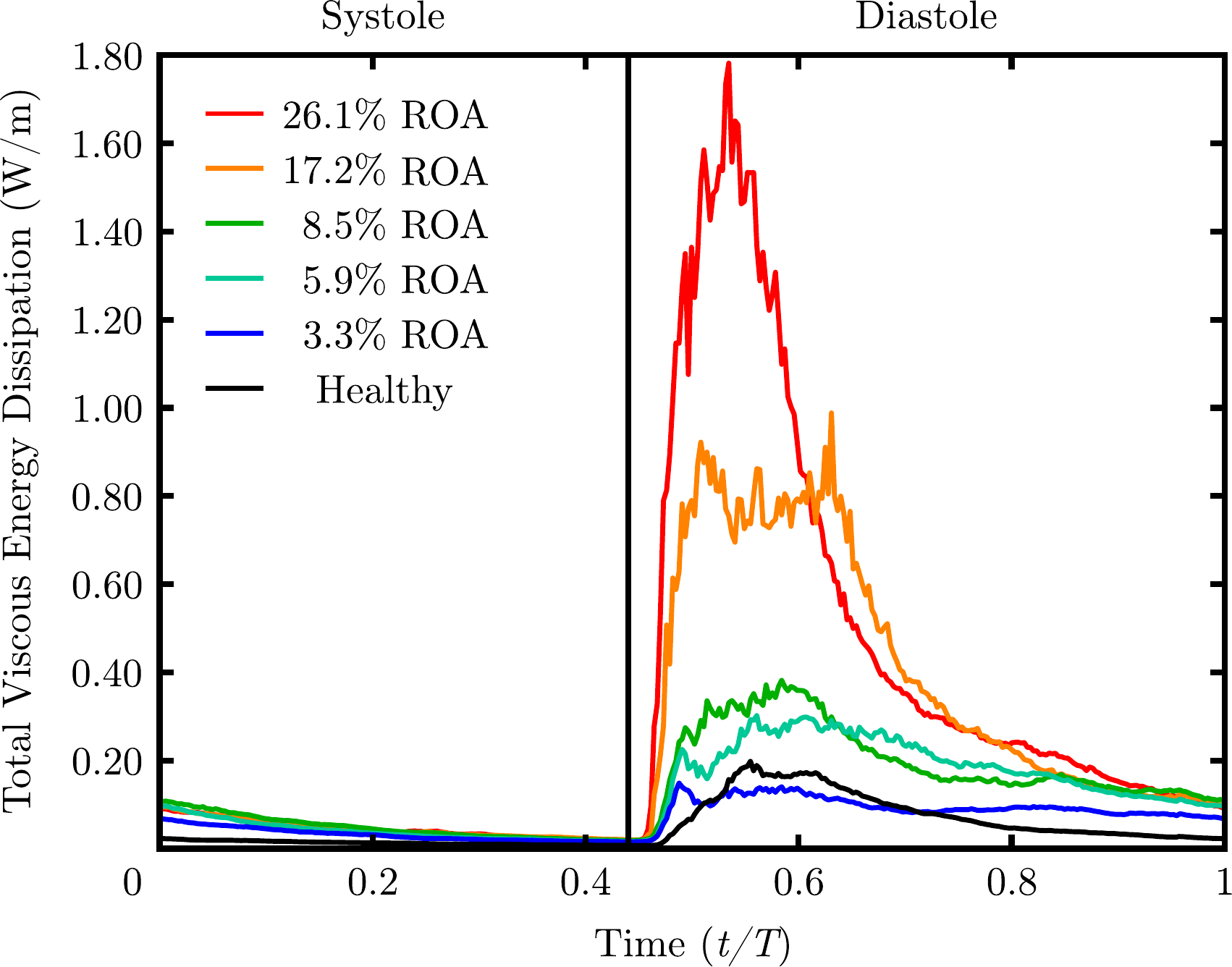}
				\caption{Evolution of viscous energy dissipation rate (W/m) over one cardiac cycle in the left ventricle for the regurgitant orifice areas ($\mathrm{ROA}$s) used in this study.}
				\label{fig:TVED}
			\end{figure}

			Following recent interest \cite{Pedrizzetti05, Stugaard15}, the rate at which energy is dissipated by viscous stresses in the flow is illustrated in Fig.\ \ref{fig:TVED}. This quantity could be seen as a measure of the efficiency of the intraventricular flow to conserve inflowing kinetic energy, suggesting a greater work input requirement from the heart to compensate for greater losses in the flow. In 2D, the total viscous energy dissipation rate ($\mathrm{VED}$) is given below (units of power per unit depth) with $\mu$ being the dynamic viscosity. Characteristically, this quantity is zero in the case of a vortex described by solid body rotation and becomes appreciable in regions of elevated longitudinal or transverse strain rate (as in jets and shear layers).
			$$
				\mathrm{VED} = \frac{\mu}{2}\int_A\left(\sum_{\forall i,j}\left(\frac{\partial u_i}{\partial x_j} + \frac{\partial u_j}{\partial x_i}\right)^2\right)\mathrm{d}A
			$$
			
\begin{figure}[!h]
				\centering
				\includegraphics[width=0.60\textwidth]{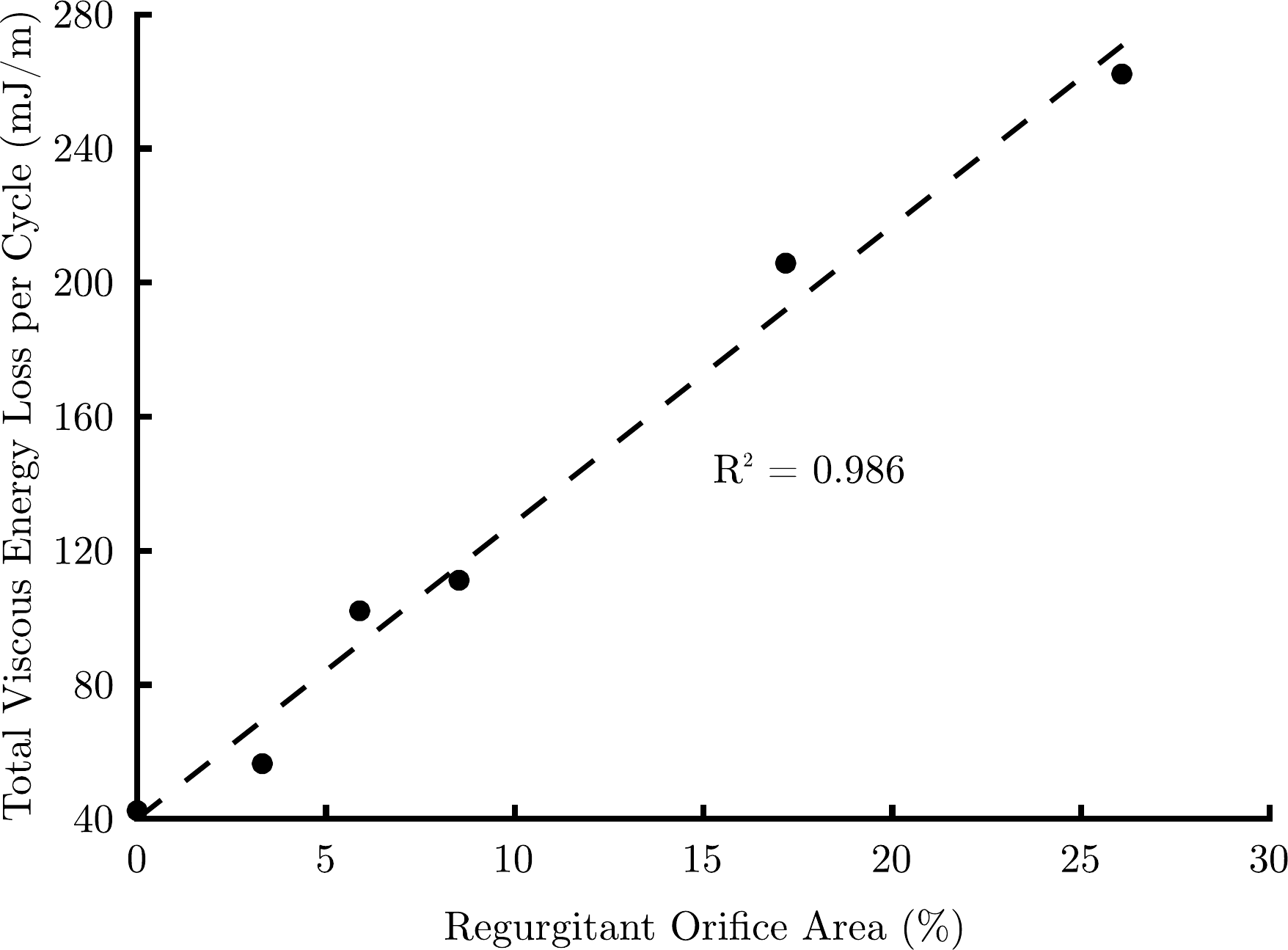}
				\caption{The total viscous energy loss per cardiac cycle (mJ/m) is seen to increase rather linearly with regurgitant orifice area.}
				\label{fig:EnergyLoss}
			\end{figure}

			A clear overall increase is observed with regurgitation severity. This is consistent with the \textit{in vivo} results reported by \citeA{Stugaard15}, despite the limitations associated with their use of vector flow mapping to acquire velocity fields \cite{PedrizzettiSengupta15}. The initial climb of the dissipation rate is associated with the high strain rates accompanying the mitral and regurgitant jets, as well as with mild turbulent fluctuations accompanying the regurgitant jet. The peak corresponds more or less to the end of the \textit{E} wave. The following decay is associated with the establishment of a vortical motion in the ventricle, whose core somewhat approximates rigid body rotation. Repeating the experiment $10$ times for each case and comparing successive curves by $p$-value calculations using two-way ANOVA, distinction between the values of total $\mathrm{VED}$ is confirmed at a $95$\% confidence level ($p_{\mathrm{max}} = 2.4 \cdot 10^{-7}$ between the healthy and mild cases). Integrating under these curves for one cycle, the energy loss per unit depth can be seen to almost linearly increase with $\mathrm{ROA}$ (Fig.\ \ref{fig:EnergyLoss}). It should be noted that although the peak $\mathrm{VED}$ in the healthy scenario is higher than that of the mild case, the energy loss is smaller owing to the prolonged effect of the perturbation imparted by the regurgitation. It appears that yet again the vortical motion in the healthy LV represents an optimal flow condition in the sense of energy conservation.
			
	\section{Conclusion}
	
		These results suggest primarily that progressive vortex reversal in the LV may naturally arise with chronic AR. The potential of the total viscous energy loss per cycle to evaluate AR severity is also highlighted, remarkably showing an almost linear relationship between energy loss and $\mathrm{ROA}$ (Fig.\ \ref{fig:EnergyLoss}), despite the overall nonlinear nature of the flow. This further suggests the energy loss to be an excellent indicator in the case of AR, being directly related to the $\mathrm{ROA}$ itself which is difficult to evaluate in practice.
	
		Evidently, there are many factors that may alter the intraventricular flow that have not been investigated here. These include ventricular geometry for instance, as well as the nature of the regurgitant orifice and jet, with angle and eccentricity being perhaps the most influential. The fluid dynamic phenomena demonstrated in this work are based on an idealized model left ventricle and it is not yet clear whether they may be generalized to all left ventricular geometries. Furthermore, while the two-dimensional plane selected for study is the plane most commonly used in clinical practice to assess AR via color Doppler, it may not be the case that the viscous energy loss exhibits such a linear monotonic increase when considering the full three-dimensional flow. These authors believe that further research is needed to consider both geometrical differences and three-dimensionality, which this group is actively working to achieve. Additionally, we would like to stress that in the case of AR, \textit{in vivo} velocity field data for a comprehensive set of severities, for which there is considerable need, will allow for better scrutiny of the aptitude of vortex and energy characteristics to assess AR severity against the current metrics.

	\section*{Acknowledgments}
	
		The authors would like to thank Drs.\ Eyal Ben-Assa, J\'{e}r\^{o}me V\'{e}tel and Marc-\'{E}tienne Lamarche-Gagnon for their constructive comments associated with this short communication, and again Dr.\ V\'{e}tel for his generosity in the use of his laboratory facilities (particularly DaVis 8.2 for this article). Additionally, the authors would like to thank the participants of the 8th International Bio-Fluid Symposium (Pasadena, CA; February, 2016) and the 69th Annual Meeting of the APS Division of Fluid Dynamics (Portland, OR; November, 2016) for their constructive comments and fruitful insight.
	
	\appendix
	\renewcommand{\theequation}{\thesection.\arabic{equation}}
	\renewcommand{\thefigure}{\thesection.\arabic{figure}}
	\renewcommand{\thetable}{\thesection.\arabic{table}}
	
	\section{\label{app:Suppl}Supplemental Material}
	\setcounter{equation}{0}
	\setcounter{figure}{0}
	\setcounter{table}{0}
	
		The reader is referred to \href{https://doi.org/10.1016/j.jbiomech.2018.07.023}{\url{https://doi.org/10.1016/j.jbiomech.2018.07.023}} for the supplemental material.
	
	\bibliographystyle{apacite}

\begin{thebibliography}{}
		
		\bibitem [\protect \citeauthoryear {%
			Adrian%
			\ \BBA {} Westerweel%
		}{%
			Adrian%
			\ \BBA {} Westerweel%
		}{%
			{\protect \APACyear {2011}}%
		}]{%
			Adrian11}
		\APACinsertmetastar {%
			Adrian11}%
		\begin{APACrefauthors}%
			Adrian, R\BPBI J.%
			\BCBT {}\ \BBA {} Westerweel, J.%
		\end{APACrefauthors}%
		\unskip\
		\newblock
		\APACrefYear{2011}.
		\newblock
		\APACrefbtitle {{Particle Image Velocimetry}} {{Particle Image Velocimetry}}.
		\newblock
		\APACaddressPublisher{New York, NY}{Cambridge University Press}.
		\PrintBackRefs{\CurrentBib}
		
		\bibitem [\protect \citeauthoryear {%
			Bekeredjian%
			\ \BBA {} Grayburn%
		}{%
			Bekeredjian%
			\ \BBA {} Grayburn%
		}{%
			{\protect \APACyear {2005}}%
		}]{%
			Bekeredjian05}
		\APACinsertmetastar {%
			Bekeredjian05}%
		\begin{APACrefauthors}%
			Bekeredjian, R.%
			\BCBT {}\ \BBA {} Grayburn, P\BPBI A.%
		\end{APACrefauthors}%
		\unskip\
		\newblock
		\APACrefYearMonthDay{2005}{}{}.
		\newblock
		{\BBOQ}\APACrefatitle
		{\href{https://doi.org/10.1161/CIRCULATIONAHA.104.488825}{Valvular heart
				disease: Aortic regurgitation}}
		{\href{https://doi.org/10.1161/CIRCULATIONAHA.104.488825}{Valvular heart
				disease: Aortic regurgitation}}.{\BBCQ}
		\newblock
		\APACjournalVolNumPages{Circulation}{112}{1}{125-134}.
		\PrintBackRefs{\CurrentBib}
		
		\bibitem [\protect \citeauthoryear {%
			{Di Labbio}%
			\ \BBA {} Kadem%
		}{%
			{Di Labbio}%
			\ \BBA {} Kadem%
		}{%
			{\protect \APACyear {2016}}%
		}]{%
			DiLabbio16}
		\APACinsertmetastar {%
			DiLabbio16}%
		\begin{APACrefauthors}%
			{Di Labbio}, G.%
			\BCBT {}\ \BBA {} Kadem, L.%
		\end{APACrefauthors}%
		\unskip\
		\newblock
		\APACrefYearMonthDay{2016}{}{}.
		\newblock
		{\BBOQ}\APACrefatitle
		{\href{http://meetings.aps.org/link/BAPS.2016.DFD.L15.2}{Vortex and energy
				characteristics of flow in the left ventricle following progressive
				severities of aortic valve regurgitation}}
		{\href{http://meetings.aps.org/link/BAPS.2016.DFD.L15.2}{Vortex and energy
				characteristics of flow in the left ventricle following progressive
				severities of aortic valve regurgitation}}.{\BBCQ}
		\newblock
		\APACjournalVolNumPages{Bulletin of the American Physical
			Society}{61}{20}{L15.2}.
		\PrintBackRefs{\CurrentBib}
		
		\bibitem [\protect \citeauthoryear {%
			Gharib%
			, Rambod%
			, Kheradvar%
			, Sahn%
			\BCBL {}\ \BBA {} Dabiri%
		}{%
			Gharib%
			\ \protect \BOthers {.}}{%
			{\protect \APACyear {2006}}%
		}]{%
			Gharib06}
		\APACinsertmetastar {%
			Gharib06}%
		\begin{APACrefauthors}%
			Gharib, M.%
			, Rambod, E.%
			, Kheradvar, A.%
			, Sahn, D\BPBI J.%
			\BCBL {}\ \BBA {} Dabiri, J\BPBI O.%
		\end{APACrefauthors}%
		\unskip\
		\newblock
		\APACrefYearMonthDay{2006}{}{}.
		\newblock
		{\BBOQ}\APACrefatitle {\href{https://doi.org/10.1073/pnas.0600520103}{Optimal
				vortex formation as an index of cardiac health}}
		{\href{https://doi.org/10.1073/pnas.0600520103}{Optimal vortex formation as
				an index of cardiac health}}.{\BBCQ}
		\newblock
		\APACjournalVolNumPages{Proceedings of the National Academy of Sciences of the
			United States of America}{103}{16}{6305-6308}.
		\PrintBackRefs{\CurrentBib}
		
		\bibitem [\protect \citeauthoryear {%
			Hendabadi%
			\ \protect \BOthers {.}}{%
			Hendabadi%
			\ \protect \BOthers {.}}{%
			{\protect \APACyear {2013}}%
		}]{%
			Hendabadi13}
		\APACinsertmetastar {%
			Hendabadi13}%
		\begin{APACrefauthors}%
			Hendabadi, S.%
			, Bermejo, J.%
			, Benito, Y.%
			, Yotti, R.%
			, Fern\'{a}ndez-Avil\'{e}s, F.%
			, {del \'{A}lamo}, J\BPBI C.%
			\BCBL {}\ \BBA {} Shadden, S\BPBI C.%
		\end{APACrefauthors}%
		\unskip\
		\newblock
		\APACrefYearMonthDay{2013}{}{}.
		\newblock
		{\BBOQ}\APACrefatitle
		{\href{https://doi.org/10.1007/s10439-013-0853-z}{Topology of blood transport
				in the human left ventricle by novel processing of Doppler echocardiography}}
		{\href{https://doi.org/10.1007/s10439-013-0853-z}{Topology of blood transport
				in the human left ventricle by novel processing of Doppler
				echocardiography}}.{\BBCQ}
		\newblock
		\APACjournalVolNumPages{Annals of Biomedical Engineering}{41}{12}{2603-2616}.
		\PrintBackRefs{\CurrentBib}
		
		\bibitem [\protect \citeauthoryear {%
			Kilner%
			\ \protect \BOthers {.}}{%
			Kilner%
			\ \protect \BOthers {.}}{%
			{\protect \APACyear {2000}}%
		}]{%
			Kilner00}
		\APACinsertmetastar {%
			Kilner00}%
		\begin{APACrefauthors}%
			Kilner, P\BPBI J.%
			, Yang, G\BPBI Z.%
			, Wilkes, A\BPBI J.%
			, Mohiaddin, R\BPBI H.%
			, Firmin, D\BPBI N.%
			\BCBL {}\ \BBA {} Yacoub, M\BPBI H.%
		\end{APACrefauthors}%
		\unskip\
		\newblock
		\APACrefYearMonthDay{2000}{}{}.
		\newblock
		{\BBOQ}\APACrefatitle {\href{https://doi.org/10.1038/35008075}{Asymmetric
				redirection of flow through the heart}}
		{\href{https://doi.org/10.1038/35008075}{Asymmetric redirection of flow
				through the heart}}.{\BBCQ}
		\newblock
		\APACjournalVolNumPages{Nature}{404}{6779}{759-761}.
		\PrintBackRefs{\CurrentBib}
		
		\bibitem [\protect \citeauthoryear {%
			Lancellotti%
			\ \protect \BOthers {.}}{%
			Lancellotti%
			\ \protect \BOthers {.}}{%
			{\protect \APACyear {2013}}%
		}]{%
			LancellottiTribouilloy13}
		\APACinsertmetastar {%
			LancellottiTribouilloy13}%
		\begin{APACrefauthors}%
			Lancellotti, P.%
			, Tribouilloy, C.%
			, Hagendorff, A.%
			, Popescu, B\BPBI A.%
			, Edvardsen, T.%
			, Pierard, L\BPBI A.%
			\BDBL {}Zamorano, J\BPBI L.%
		\end{APACrefauthors}%
		\unskip\
		\newblock
		\APACrefYearMonthDay{2013}{}{}.
		\newblock
		{\BBOQ}\APACrefatitle
		{\href{https://doi.org/10.1093/ehjci/jet105}{Recommendations for the
				echocardiographic assessment of native valvular regurgitation: An executive
				summary from the European Association of Cardiovascular Imaging}}
		{\href{https://doi.org/10.1093/ehjci/jet105}{Recommendations for the
				echocardiographic assessment of native valvular regurgitation: An executive
				summary from the European Association of Cardiovascular Imaging}}.{\BBCQ}
		\newblock
		\APACjournalVolNumPages{European Heart Journal -- Cardiovascular
			Imaging}{14}{7}{611-644}.
		\PrintBackRefs{\CurrentBib}
		
		\bibitem [\protect \citeauthoryear {%
			Narula%
			, Vannan%
			\BCBL {}\ \BBA {} DeMaria%
		}{%
			Narula%
			\ \protect \BOthers {.}}{%
			{\protect \APACyear {2007}}%
		}]{%
			Narula07}
		\APACinsertmetastar {%
			Narula07}%
		\begin{APACrefauthors}%
			Narula, J.%
			, Vannan, M\BPBI A.%
			\BCBL {}\ \BBA {} DeMaria, A\BPBI N.%
		\end{APACrefauthors}%
		\unskip\
		\newblock
		\APACrefYearMonthDay{2007}{}{}.
		\newblock
		{\BBOQ}\APACrefatitle {\href{https://doi.org/10.1016/j.jacc.2006.12.006}{Of
				that Waltz in my heart}}
		{\href{https://doi.org/10.1016/j.jacc.2006.12.006}{Of that Waltz in my
				heart}}.{\BBCQ}
		\newblock
		\APACjournalVolNumPages{Journal of the American College of
			Cardiology}{49}{8}{917-920}.
		\PrintBackRefs{\CurrentBib}
		
		\bibitem [\protect \citeauthoryear {%
			Nishimura%
			\ \protect \BOthers {.}}{%
			Nishimura%
			\ \protect \BOthers {.}}{%
			{\protect \APACyear {2014}}%
		}]{%
			Nishimura14}
		\APACinsertmetastar {%
			Nishimura14}%
		\begin{APACrefauthors}%
			Nishimura, R\BPBI A.%
			, Otto, C\BPBI M.%
			, Bonow, R\BPBI O.%
			, Carabello, B\BPBI A.%
			, {Erwin III}, J\BPBI P.%
			, Guyton, R\BPBI A.%
			\BDBL {}Thomas, J\BPBI D.%
		\end{APACrefauthors}%
		\unskip\
		\newblock
		\APACrefYearMonthDay{2014}{}{}.
		\newblock
		{\BBOQ}\APACrefatitle {\href{https://doi.org/10.1016/j.jacc.2014.02.536}{2014
				AHA/ACC guideline for the management of patients with valvular heart disease:
				A report of the American College of Cardiology/American Heart Association
				Task Force on Practice Guidelines}}
		{\href{https://doi.org/10.1016/j.jacc.2014.02.536}{2014 AHA/ACC guideline for
				the management of patients with valvular heart disease: A report of the
				American College of Cardiology/American Heart Association Task Force on
				Practice Guidelines}}.{\BBCQ}
		\newblock
		\APACjournalVolNumPages{Journal of the American College of
			Cardiology}{63}{22}{e57-e185}.
		\PrintBackRefs{\CurrentBib}
		
		\bibitem [\protect \citeauthoryear {%
			Pedrizzetti%
			\ \BBA {} Domenichini%
		}{%
			Pedrizzetti%
			\ \BBA {} Domenichini%
		}{%
			{\protect \APACyear {2005}}%
		}]{%
			Pedrizzetti05}
		\APACinsertmetastar {%
			Pedrizzetti05}%
		\begin{APACrefauthors}%
			Pedrizzetti, G.%
			\BCBT {}\ \BBA {} Domenichini, F.%
		\end{APACrefauthors}%
		\unskip\
		\newblock
		\APACrefYearMonthDay{2005}{}{}.
		\newblock
		{\BBOQ}\APACrefatitle
		{\href{https://doi.org/10.1103/PhysRevLett.95.108101}{Nature optimizes the
				swirling flow in the human left ventricle}}
		{\href{https://doi.org/10.1103/PhysRevLett.95.108101}{Nature optimizes the
				swirling flow in the human left ventricle}}.{\BBCQ}
		\newblock
		\APACjournalVolNumPages{Physical Review Letters}{95}{10}{108101}.
		\PrintBackRefs{\CurrentBib}
		
		\bibitem [\protect \citeauthoryear {%
			Pedrizzetti%
			\ \BBA {} Sengupta%
		}{%
			Pedrizzetti%
			\ \BBA {} Sengupta%
		}{%
			{\protect \APACyear {2015}}%
		}]{%
			PedrizzettiSengupta15}
		\APACinsertmetastar {%
			PedrizzettiSengupta15}%
		\begin{APACrefauthors}%
			Pedrizzetti, G.%
			\BCBT {}\ \BBA {} Sengupta, P\BPBI P.%
		\end{APACrefauthors}%
		\unskip\
		\newblock
		\APACrefYearMonthDay{2015}{}{}.
		\newblock
		{\BBOQ}\APACrefatitle {\href{https://doi.org/10.1093/ehjci/jev070}{Vortex
				imaging: New information gain from tracking cardiac energy loss}}
		{\href{https://doi.org/10.1093/ehjci/jev070}{Vortex imaging: New information
				gain from tracking cardiac energy loss}}.{\BBCQ}
		\newblock
		\APACjournalVolNumPages{European Heart Journal -- Cardiovascular
			Imaging}{16}{7}{719-720}.
		\PrintBackRefs{\CurrentBib}
		
		\bibitem [\protect \citeauthoryear {%
			Raffel%
			, Willert%
			, Wereley%
			\BCBL {}\ \BBA {} Kompenhans%
		}{%
			Raffel%
			\ \protect \BOthers {.}}{%
			{\protect \APACyear {2007}}%
		}]{%
			Raffel07}
		\APACinsertmetastar {%
			Raffel07}%
		\begin{APACrefauthors}%
			Raffel, M.%
			, Willert, C.%
			, Wereley, S\BPBI T.%
			\BCBL {}\ \BBA {} Kompenhans, J.%
		\end{APACrefauthors}%
		\unskip\
		\newblock
		\APACrefYear{2007}.
		\newblock
		\APACrefbtitle {\href{https://doi.org/10.1007/978-3-540-72308-0}{Particle Image
				Velocimetry: A Practical Guide}}
		{\href{https://doi.org/10.1007/978-3-540-72308-0}{Particle Image Velocimetry:
				A Practical Guide}}\ (\PrintOrdinal{2}\ \BEd).
		\newblock
		\APACaddressPublisher{New York, NY}{Springer}.
		\PrintBackRefs{\CurrentBib}
		
		\bibitem [\protect \citeauthoryear {%
			Raymondet%
			, Kadem%
			\BCBL {}\ \BBA {} {Di Labbio}%
		}{%
			Raymondet%
			\ \protect \BOthers {.}}{%
			{\protect \APACyear {2016}}%
		}]{%
			Raymondet16}
		\APACinsertmetastar {%
			Raymondet16}%
		\begin{APACrefauthors}%
			Raymondet, A.%
			, Kadem, L.%
			\BCBL {}\ \BBA {} {Di Labbio}, G.%
		\end{APACrefauthors}%
		\unskip\
		\newblock
		\APACrefYearMonthDay{2016}{}{}.
		\newblock
		{\BBOQ}\APACrefatitle {{Jet-vortex interaction in the left ventricle during
				diastole in the presence of aortic regurgitation}} {{Jet-vortex interaction
				in the left ventricle during diastole in the presence of aortic
				regurgitation}}.{\BBCQ}
		\newblock
		\BIn{} \APACrefbtitle {{Proceedings of the Eighth International Bio-Fluid
				Symposium}.} {{Proceedings of the Eighth International Bio-Fluid Symposium}.}
		\newblock
		\APACaddressPublisher{Pasadena, CA}{}.
		\PrintBackRefs{\CurrentBib}
		
		\bibitem [\protect \citeauthoryear {%
			Singh%
			\ \protect \BOthers {.}}{%
			Singh%
			\ \protect \BOthers {.}}{%
			{\protect \APACyear {1999}}%
		}]{%
			Singh99}
		\APACinsertmetastar {%
			Singh99}%
		\begin{APACrefauthors}%
			Singh, J\BPBI P.%
			, Evans, J\BPBI C.%
			, Levy, D.%
			, Larson, M\BPBI G.%
			, Freed, L\BPBI A.%
			, Fuller, D\BPBI L.%
			\BDBL {}Benjamin, E\BPBI J.%
		\end{APACrefauthors}%
		\unskip\
		\newblock
		\APACrefYearMonthDay{1999}{}{}.
		\newblock
		{\BBOQ}\APACrefatitle
		{\href{https://doi.org/10.1016/S0002-9149(98)01064-9}{Prevalence and clinical
				determinants of mitral, tricuspid, and aortic regurgitation (the Framingham
				Heart Study)}}
		{\href{https://doi.org/10.1016/S0002-9149(98)01064-9}{Prevalence and clinical
				determinants of mitral, tricuspid, and aortic regurgitation (the Framingham
				Heart Study)}}.{\BBCQ}
		\newblock
		\APACjournalVolNumPages{The American Journal of Cardiology}{83}{6}{897-902}.
		\PrintBackRefs{\CurrentBib}
		
		\bibitem [\protect \citeauthoryear {%
			Stout%
			\ \BBA {} Verrier%
		}{%
			Stout%
			\ \BBA {} Verrier%
		}{%
			{\protect \APACyear {2009}}%
		}]{%
			Stout09}
		\APACinsertmetastar {%
			Stout09}%
		\begin{APACrefauthors}%
			Stout, K\BPBI K.%
			\BCBT {}\ \BBA {} Verrier, E\BPBI D.%
		\end{APACrefauthors}%
		\unskip\
		\newblock
		\APACrefYearMonthDay{2009}{}{}.
		\newblock
		{\BBOQ}\APACrefatitle
		{\href{https://doi.org/10.1161/CIRCULATIONAHA.108.782292}{Acute valvular
				regurgitation}}
		{\href{https://doi.org/10.1161/CIRCULATIONAHA.108.782292}{Acute valvular
				regurgitation}}.{\BBCQ}
		\newblock
		\APACjournalVolNumPages{Circulation}{119}{25}{3232-3241}.
		\PrintBackRefs{\CurrentBib}
		
		\bibitem [\protect \citeauthoryear {%
			Stugaard%
			\ \protect \BOthers {.}}{%
			Stugaard%
			\ \protect \BOthers {.}}{%
			{\protect \APACyear {2015}}%
		}]{%
			Stugaard15}
		\APACinsertmetastar {%
			Stugaard15}%
		\begin{APACrefauthors}%
			Stugaard, M.%
			, Koriyama, H.%
			, Katsuki, K.%
			, Masuda, K.%
			, Asanuma, T.%
			, Takeda, Y.%
			\BDBL {}Nakatani, S.%
		\end{APACrefauthors}%
		\unskip\
		\newblock
		\APACrefYearMonthDay{2015}{}{}.
		\newblock
		{\BBOQ}\APACrefatitle {\href{https://doi.org/10.1093/ehjci/jev035}{Energy loss
				in the left ventricle obtained by vector flow mapping as a new quantitative
				measure of severity of aortic regurgitation: A combined experimental and
				clinical study}} {\href{https://doi.org/10.1093/ehjci/jev035}{Energy loss in
				the left ventricle obtained by vector flow mapping as a new quantitative
				measure of severity of aortic regurgitation: A combined experimental and
				clinical study}}.{\BBCQ}
		\newblock
		\APACjournalVolNumPages{European Heart Journal -- Cardiovascular
			Imaging}{16}{7}{723-730}.
		\PrintBackRefs{\CurrentBib}
		
	\end{thebibliography}

\end{document}